\documentclass[10pt,letterpaper,twocolumn]{article} 

\usepackage{ol}

\begin{document}

\twocolumn[ 

\title{Generation of a squeezed vacuum resonant on Rubidium D$_1$ line \\ with periodically-poled KTiOPO$\mathrm{_4}$}

\author{Takahito Tanimura, Daisuke Akamatsu, Yoshihiko Yokoi}
\address{Department of Physics, Tokyo Institute of Technology, 2-12-1 O-okayama, Meguro-ku, Tokyo 152-8550, Japan}

\author{Akira Furusawa}
\address{Department of Applied Physics, School of Engineering, The University of Tokyo, \\7-3-1 Hongo, Bunkyo-ku, Tokyo 113-8656, Japan}

\author{Mikio Kozuma}
\address{Department of Physics, Tokyo Institute of Technology, 2-12-1 O-okayama, Meguro-ku, Tokyo 152-8550, Japan \\
PRESTO, Japan Science and Technology Agency, 4-1-8 Honcho Kawaguchi, Saitama, Japan \\
CREST, Japan Science and Technology Agency, 1-9-9 Yaesu, Chuo-ku, Tokyo 103-0028, Japan}


\begin{abstract}We report generation of a continuous-wave squeezed vacuum resonant on the Rb D$_1$ line (795 nm) using periodically poled KTiOPO$_4$ (PPKTP) crystals. 
With a frequency doubler and an optical parametric oscillator based on PPKTP crystals, we observed a squeezing level of $-2.75 \pm 0.14 \mathrm{dB}$ and an anti-squeezing level of $+7.00 \pm 0.13 \mathrm{dB}$. This system could be utilized for demonstrating storage and retrieval of the squeezed vacuum, which is important for the ultra-precise measurement of atomic spins as well as quantum information processing.
\end{abstract} 


 ] 

\noindent 
Recently, a novel scheme for mapping the quantum state of a light field onto an atomic ensemble was proposed \cite{EITPRA} in which the electromagnetically induced transparency plays a major role. This ``storage of light'' technique enables us to overcome the difficulty of localizing photons which are mainly used as carriers of quantum information. While the storage and retrieval of a single photon state has already been realized, \cite{DLCZ-EIT, Kuz} it has not been demonstrated for a squeezed vacuum. It should be noted that the former experiment can be performed conditionally whereas the latter should be demonstrated in deterministic manner and is thus sensitive to field loss. Mapping the squeezed state onto an atomic ensemble is a critical task not only for quantum information processing but also for ultra-precise measurement of atomic spins. 

To perform such an experiment, it is necessary to generate a high-level squeezed vacuum resonant on an atomic transition. By utilizing KNbO$_3$ crystals, a squeezed vacuum has already been generated resonant on the Cs D-lines (852 nm, 894 nm) and the interaction between the atoms and the squeezed vacuum has also been investigated intensively  \cite{ESP92}. However, there have been relatively few experiments done on the generation of a squeezed vacuum resonant on the Rb D-lines (780 nm, 795 nm), while Rb has played an important role in quantum information processing along with Cs. So far experiments have been limited to -0.9 dB squeezing with quasi-phase-matched MgO:LiNbO$_3$ waveguides \cite{AKA-PRL} and -0.85 dB squeezing with the self-rotation of Rb itself \cite{self-rotation}. Note that the KNbO$_3$ crystal which is useful at the Cs resonance line cannot be utilized at the Rb one \cite{F1}.  In this letter we demonstrate -2.75 dB squeezing at 795 nm using a periodically-poled KTiOPO$_4$ (PPKTP) crystal \cite{F1, F2}, which to the best of our knowledge, is the highest squeezing obtained at the Rb D$_1$ line. 

\begin{figure}[htb]
\centerline{\includegraphics[]{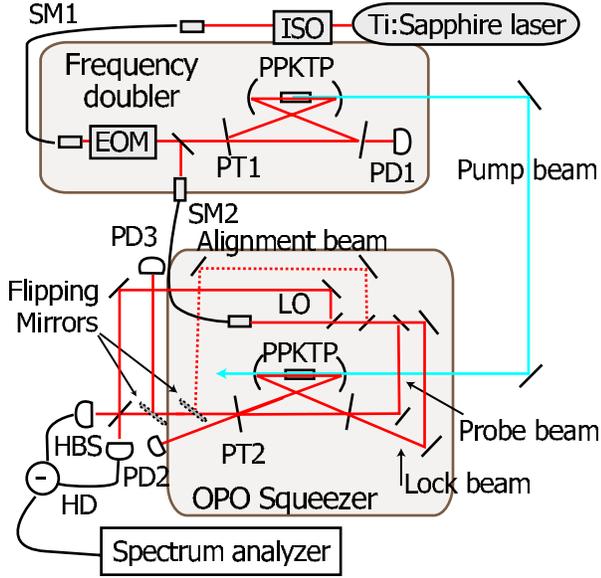}}
\caption{Experimental setup. ISO: optical isolator, EOM: electro optic modulator, OPO: sub-threshold degenerate optical parametric oscillator, HBS: 50:50 beam splitter, PT: partial transmittance mirror, HD: balanced homodyne detector, PD: photodiode, SM: single mode fiber.}
\label{Abst_E_setup}
\end{figure}

\begin{figure}[htb]
\centerline{\includegraphics[]{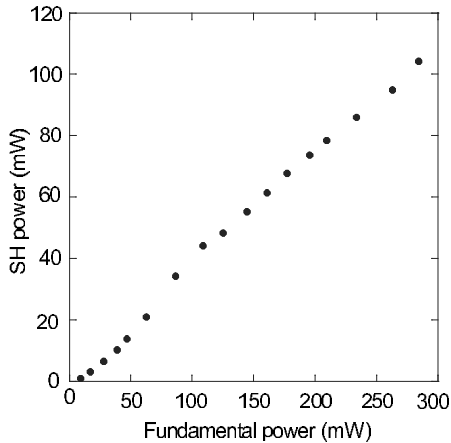}}
\caption{ Second-harmonic output power versus incident laser power.}
\label{b}
\end{figure}

Figure \ref{Abst_E_setup} shows the experimental setup.
A continuous-wave Ti:Sapphire laser (Coherent, MBR 110) at 795 nm was employed in this experiment.
The beam from the Ti:Sapphire laser was phase-modulated by an electro-optic modulator (EOM).
This modulation was utilized to lock a cavity for frequency doubling and a cavity for squeezing using the FM sideband method \cite{Drever83}.
The frequency doubler had a bow-tie type ring configuration with two spherical mirrors (radius of curvature of 100 mm) and two flat mirrors.
One of the flat mirrors (PT1) had a reflectivity of $90 \%$ at 795 nm and was used as the input coupler, while the other mirrors were high-reflectivity coated.
All the mirrors had reflectivities of less than $5 \%$ at 397.5 nm.
A 10 mm long PPKTP crystal (Raicol Crystals) was used for second harmonic generation.
Figure \ref{b} shows the dependence of the blue output power on the fundamental laser power incident on the frequency doubler.
The output from the frequency doubler was stable over tens of minutes provided the fundamental power was less than 290 mW.

The generated 397.5 nm beam pumped a degenerate optical parametric oscillator (OPO) that also had a bow-tie type ring configuration using two spherical mirrors (radius of curvature of 50 mm) and two flat mirrors.
One of the flat mirrors (PT2) had a reflectivity of $90 \%$ at 795 nm and was used as the output coupler.
The round trip cavity length was $l = 600 mm$ and a beam waist radius inside the crystal was $20$ $\mu m$.
A 10 mm long PPKTP crystal was again used for parametric down conversion.
The OPO was driven below the parametric oscillation threshold $P_{th} = 150 \mathrm{mW}$, which was derived theoretically from the nonlinear efficiency of the crystal $E_{NL} = 0.023 \mathrm{W^{-1}}$, the intra-cavity loss $L = 0.0173$, and the transmittance of the input coupler $T = 0.10$ using the following expression 
$
P_{th} = (T + L)^2/(4 E_{NL}).
$

The IR beam from the OPO squeezer was split into four beams: an alignment beam, a probe beam, a lock beam, and a local oscillator beam for homodyne detection. The alignment beam was an auxiliary beam and was used for aligning the cavity, measuring the intra-cavity loss, and matching the spatial mode of the pump beam with the OPO cavity.
For the last application,
the alignment beam was converted to the second harmonic using the OPO and used as a reference beam.
This reference second harmonic beam propagated in the opposite direction to the pump beam and represented the OPO cavity mode.
This meant that by matching the spatial mode of the reference beam with that of the pump beam, the pump beam could be matched with the OPO cavity mode.

The probe beam was utilized to measure the classical parametric gain.
This was done by injecting the beam into the OPO cavity through a high-reflection flat mirror and detecting the transmitted probe beam from the output coupler with a photodiode (PD3).

The lock beam was also injected into the cavity through a high-reflection flat mirror in the counter propagating mode to the probe beam.
The transmitted lock beam from the output coupler was detected with a photodiode (PD2) and the error signal for locking the cavity length was extracted using the FM sideband method.

The generated squeezed light was combined with a local oscillator (LO) at a half-beam splitter (HBS) and detected by a balanced homodyne detector (HD).
The HD had two photodiodes (Hamamatsu photonics, S-3590 with anti-reflection coating) that had a quantum efficiency of $98 \% $.
The output of the HD was measured at the sideband component of $1 \mathrm{MHz}$ using a spectrum analyzer.
The circuit noise level of the homodyne detector was 14.0 dB below the shot noise level.

\begin{figure}[htb]
\centerline{\includegraphics[]{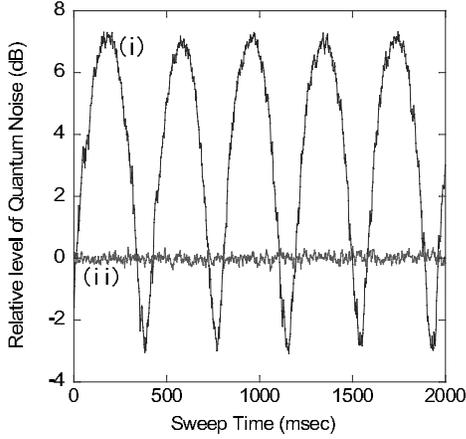}}
\caption{Measured quantum noise levels.(i) Local oscillator beam phase was scanned.(ii) Shot noise level. Noise levels are displayed as the relative power compared to the shot noise level (0 dB). The settings of the spectrum analyzer were, zero-span mode at 1 MHz, resolution bandwidth = 100 kHz, video bandwidth = 30 Hz.}
\label{Abst_E_sv_exp}
\end{figure}

Figure \ref{Abst_E_sv_exp} shows the measured quantum noise levels at a pump power of 61 mW as the local oscillator phase was scanned.
The noise level was measured with a spectrum analyzer in zero-span mode at 1 MHz, with a resolution bandwidth of 100 kHz and a video bandwidth of 30 Hz.
The squeezing level of $-2.75 \pm 0.14 \mathrm{dB}$ and the anti-squeezing level of $+7.00 \pm 0.13 \mathrm{dB}$ were observed, where the standard deviation was estimated from a fitting based on eq. (\ref{Abst_E_sqv}).

The variance of the output mode $S$ can be modeled using \cite{ESP92, Collett84}
\begin{eqnarray}
S = 1 + 4 \alpha \rho x \left[ \frac{\cos^2 \theta}{(1 - x)^2 + 4 \Omega^2} - \frac{\sin^2 \theta}{(1 + x)^2 + 4 \Omega^2} \right], \label{Abst_E_sqv}
\end{eqnarray}
where $\theta$ is the relative phase between the squeezed light and LO, $\alpha$ and $\rho$ are the detection efficiency and the OPO escape efficiency, respectively.
The detection efficiency $\alpha$ is the product of the photodiode quantum efficiency $\eta$, and the homodyne efficiency $\xi^2$ (where $\xi$ is the visibility between the output and the local oscillator mode): $\alpha = \eta \xi^2$.
The OPO escape efficiency can be written as 
$
\rho = T/(T+L),
$
where $T$ and $L$ are the transmission of the output coupler and the intra-cavity loss, respectively.
The pump parameter $x$ is defined by the pump power $P_{pump}$ and the oscillation threshold $P_{th}$, and is expressed using the parametric gain $G$ by
$
x  \equiv   (P_{pump} / P_{th})^{1/2}
  =  1 - 1/G^{1/2}.
$
The detuning parameter $\Omega$ is given by the ratio of the measurement frequency $\omega$ to the OPO cavity decay rate $\gamma = c(T+L)/l$, i.e.  $\Omega = \omega / \gamma$, where $c$ is the speed of light.

In our setup, $\eta = 0.99$ and $\xi = 0.91$, therefore $\alpha = 0.82$.
$T = 0.10$ and $L = 0.0173$ yield $\rho = 0.85$.
Note that our crystal made no measurable difference to the intra-cavity loss in the presence of the pump beam \cite{F1}. The detuning parameter was $\Omega = 0.107$.
The classical parametric gain of $G = 5.3$ measured with a 61 mW pump light yields $x = 0.57$.
With these values, eq. (\ref{Abst_E_sqv}), predicts theoretical squeezing / anti-squeezing levels of $-4.4 \mathrm{dB}$ and $+8.9 \mathrm{dB}$, respectively.
This theoretical squeezing / anti-squeezing levels are become to $-4.1 \mathrm{dB}$ and $+8.7 \mathrm{dB}$ by accounting for the effect of the circuit noise.

\begin{figure}[htb]
\centerline{\includegraphics[]{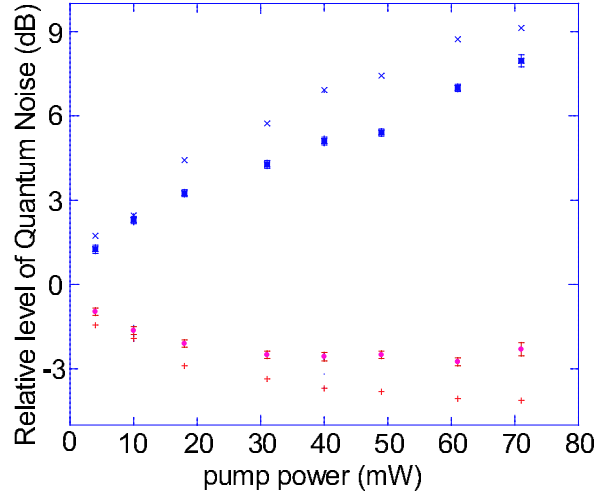}}
\caption{Squeezing and anti-squeezing levels for several powers of the pump beam. The circles and squares indicate measured values, while $+$ and $\times$ indicate theoretical values which are calculated using the parametric gains.}
\label{d}
\end{figure}

We repeated the above measurement and analysis for various pump powers.
The results are summarized in Fig. \ref{d}. There is a similar discrepancy from the theoretical values for both squeezing and anti-squeezing data. This discrepancy can not be explained by simply introducing unknown field loss, because a squeezing is theoretically much sensitive to the field loss compared to an anti-squeezing. In the current setup, the spatial mode of the incident probe beam was not perfectly matched to that of the OPO cavity. Therefore, thermal lens effect caused by injection of the blue pump beam could modify the cavity mode and increase (or decrease) the transmittance of the probe beam. When we measured the parametric gain $G$, we set the transmittance of the probe beam without the pump beam to unity. If the thermal lens effect discussed above had occurred, measured parametric gain should be corrected. Introducing about 10$\%$ correction for the value of $G$ and considering unknown field loss, the discrepancy between theoretical values and squeezing/anti-squeezing data can be explained. The observed squeezing level became slightly degraded when the pump power reached 70 mW, which could be explained by mixing of the highly anti-squeezed component with the observed quadrature noise due to the temporal fluctuation in the LO phase. 

In conclusion, we observed $-2.75 \pm 0.14 \mathrm{dB}$ squeezing and $+7.00 \pm 0.13 \mathrm{dB}$ anti-squeezing at 795 nm, which corresponds to the D$_1$ transition of Rb atoms. It should be possible to achieve a higher squeezing level by increasing visibility of the homodyne system and reducing the phase fluctuation by stabilizing the setup actively. While electromagnetically induced transparency was observed with the squeezed vacuum in our previous work \cite{AKA-PRL}, neither slow propagation nor storage could be realized due mainly to the low squeezing level. The squeezing level obtained in this study was much higher than that previously obtained with the PPLN waveguide and we thus believe that storage of the squeezed vacuum should be achievable with the current setup. 

 We gratefully acknowledge G. Takahashi, N. Takei, H. Yonezawa, and K. Wakui for their valuable comments and stimulating discussions. This work was supported by Grant-in-Aid for Young Scientists (A), and a 21st Century COE Program at Tokyo Tech ``Nanometer-Scale Quantum Physics'' by MEXT.


\end{document}